\documentstyle[12pt,epsf,fleqn]{elsart}

\def \ni{\noindent}

\def \be {\begin{equation}}
\def \ee {\end{equation}}

\begin{document}

\begin{frontmatter}
\title{High-purity germanium detector ionization pulse
shapes of nuclear recoils, -$\gamma$ interactions and microphonism}
\ni {\bf Authors} 

\author{L.~Baudis, J. Hellmig, H.V. Klapdor--Kleingrothaus, Y. Ramachers}
\address{Max--Planck--Institut f\"ur Kernphysik, Heidelberg, Germany}
\author{J. W. Hammer, A. Mayer}
\address{Institut f\"ur Strahlenphysik, Stuttgart, Germany}

\begin{abstract}
Nuclear recoil measurements with high--purity Germanium detectors are
very promising to directly detect dark matter candidates. The
main background sources in such experiments are natural radioactivity and
microphonic noise. Digital pulse shape analysis is an encouraging approach to 
reduce the
background originating from the latter. To study the pulse shapes of
nuclear recoil events we performed a neutron scattering experiment,
which covered the ionization energy range from 20 to 80 keV. We have
measured ionization efficiencies as well and found  an excellent
agreement with the theory of Lindhard. In a further experiment we
measured pulse shapes of a radioactive $\gamma$--source and found
no difference to nuclear recoil pulse shapes. Pulse shapes originating 
from microphonics of a HPGe--detector are presented for the first
time. A microphonic noise
suppression method, crucial for dark matter direct detection
experiments, can therefore be calibrated with pulse
shapes from $\gamma$--sources.
\end{abstract}
\end{frontmatter}

\section{Introduction}

The dark matter in the Galactic halo is assumed to be dominantly
composed of  WIMPs \cite{kam96}. A direct detection method is 
through WIMP interaction
with ordinary matter by elastic scattering off nuclei \cite{smi90}. Direct
detection experiments search for the energy  deposition produced in a low
background detector by a WIMP elastically scattered off a nucleus
therein (typical below 100 keV).
Most promising future approaches include experiments with
scintillation-- \cite{spo94}, cryogenic-- \cite{shu92} and semiconductor
detectors \cite{bec94}. The best results at present
are obtained using NaI--crystals \cite{spo94,ber96} and HPGe--detectors \cite{bec94}. 
The background of HPGe--detectors in the energy region below 100 keV
originates from natural radioactivity and microphonic noise. 
Therefore a further step
in reduction of the background from natural radioactivity by one order
of magnitude or more \cite{bau97} needs a reliable method to identify
microphonic noise. The success of digital pulse shape analysis \cite{hei97} in
discriminating single and multiple scattered events is encouraging
to think of a similar method in order to identify microphonics in the low
energy region of HPGe--detectors. 
Another way to identify microphonics is the simultaneous use of two
different shaping times in the processing of the signal (see
\cite{garcia95} and references therein) which is however not the topic 
of this paper.
Also are  there other applications which use
the information from the pulse shapes of the charge current of
Ge--detectors \cite{bam91,asp94,rot84}. All of the cited papers measure pulse
shapes at higher energies (beyond 200keV).  

A first step in developing a pulse shape analysis method  is to study
pulses of well defined
origin. Nuclear recoil events comparable to those of not yet known 
particles, WIMPs, can be generated by elastically scattering of neutrons off
nuclei in the germanium detector. Gamma interaction events in the
energy region below 100keV can be generated for example by radiation
of the detector with a
$^{133}$Ba source.

We measured the pulse shape of neutron interactions, of
$\gamma$--interactions and of microphonic noise. Simultaneously we measured
the ionization efficiency of germanium recoil nuclei inside germanium.
Nuclear recoil events were studied until now by a cryogenic
experiment \cite{shu92}, which could demonstrate the difference 
in ionization and phonon signals produced by
nuclear recoil and photon interactions.
The ionization efficiency
 was measured already in the 60s.
With exception of one
early experiment, which measured the endpoint of the energy spectrum from
elastically scattered neutrons \cite{sat65}, 
the shapes of a peak from inelastically
scattered neutrons \cite{cha65,cha68,mes95,jon71,jon75} were studied. 

From the good agreement of our measured ionization efficiency with the
theory of Lindhard \cite{lin63} and the previous measurements we
conclude to measure  indeed pulse shapes of Ge recoil events in
germanium.
To sample the pulse shape of each recoil event we had to perform an
event-by-event measurement.
For this purpose we built up  a coincidence experiment as described in
section 2. In section 3 we discuss the measurement of $\gamma$--ray
pulse shapes and in section 4 the measurement of microphonic pulses.
We give a conclusion and an outlook in section 5.

\section{Neutron Scattering Experiment}

\begin{figure}[h]
\epsfxsize=12cm
\epsffile{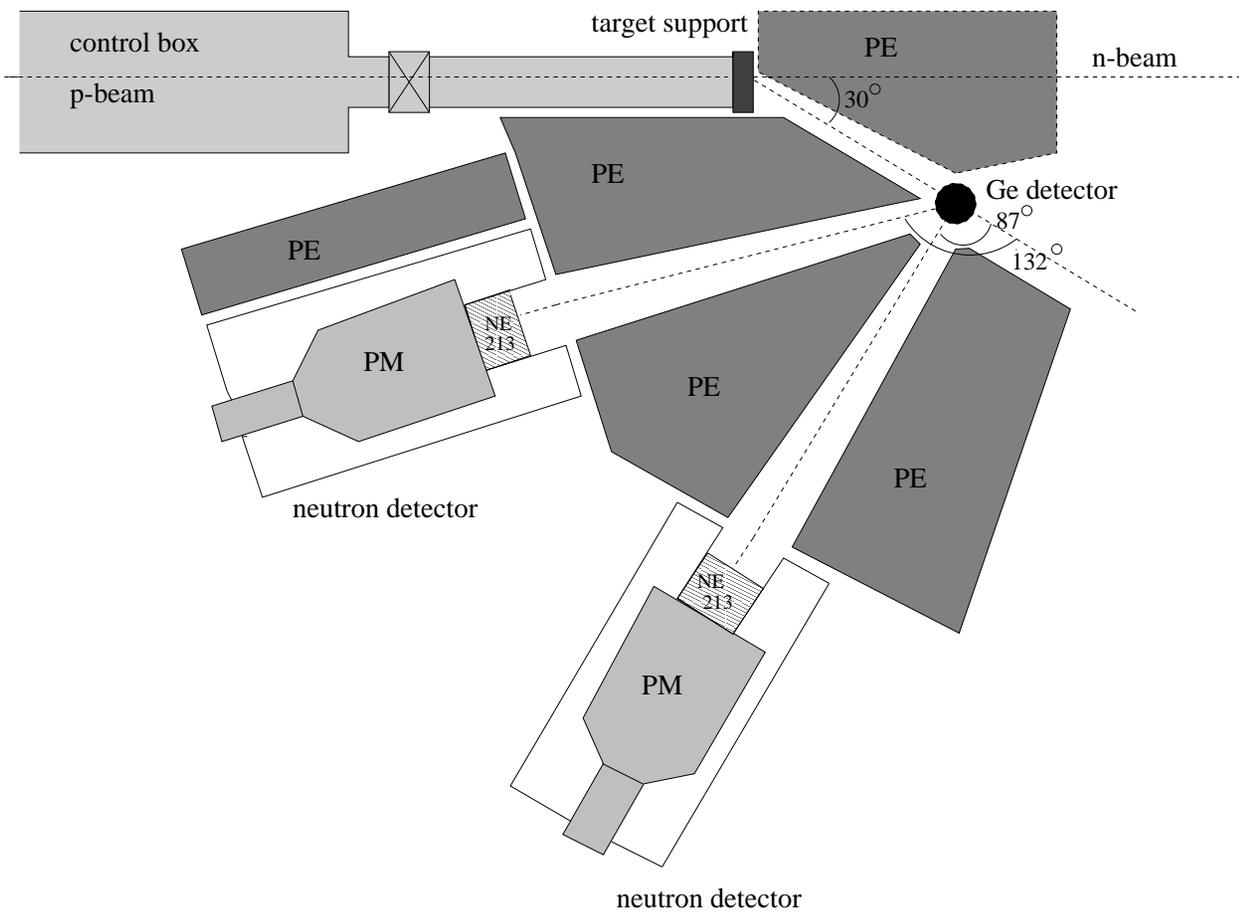}
\caption{The experimental setup for production of nuclear recoils inside a
HPGe detector.}\label{exp}
\end{figure}

The experimental setup  can be seen in Fig.~\ref{exp}.
A 3.3 MHz pulsed proton beam with 16 MeV energy, 1 ns duration, 
and 1.5 nA current was used to produce neutrons in a 
Lithium coated copper target by p($^7$Li,$^7$Be)n,
p($^{65}$Cu,$^{65}$Zn)n and p($^{63}$Cu,$^{63}$Zn)n reactions.
The maximal neutron energies for E$_{p}$ = 16 MeV under 30$^{\circ}$ are
listed in Tab.~\ref{t1}.
Due to the different reactions and the large number of excited states
in $^{65}$Cu the neutrons had a continuous energy spectrum
which was measured in 1.36 m distance by time of flight (TOF).

To select events from elastic scattering of neutrons inside
the germanium we placed NE 213 scintillators under 87
and 132 degrees (compare Fig.~1). 
The scintillators were equipped with n,$\gamma$--discrimination
using the differences in pulse rise times of neutron and 
$\gamma$--interactions in the liquid scintillator \cite{ham86}.
Coincidence between the timing signal of the Ge--detector, 
one of the scintillators and of the proton beam signal was used as
start signal for the time measurements. A coincidence of
the delayed start signal and the n,$\gamma$--discrimination reduced the
trigger rate down to 1 Hz due to rejection of random coincidences of
$\gamma$--interactions in the scintillators and the Ge--detector.

For each event we recorded energy--deposits inside the HPGe and the
neutron detectors, n,$\gamma$--signal of the neutron detectors, time
differences between beam pulse and each of the three detectors and the
pulse shape of the differentiated HPGe preamplifier output.

Our Ge-detector was a n-type coaxial, closed ended HPGe detector with
a mass of 1.05 kg and a diameter of 5cm. N-type
detectors are more resistant to fast neutron damage than p--types 
\cite{kra80,esc94}.

\begin{figure}[h]
\epsfxsize=15cm
\epsffile{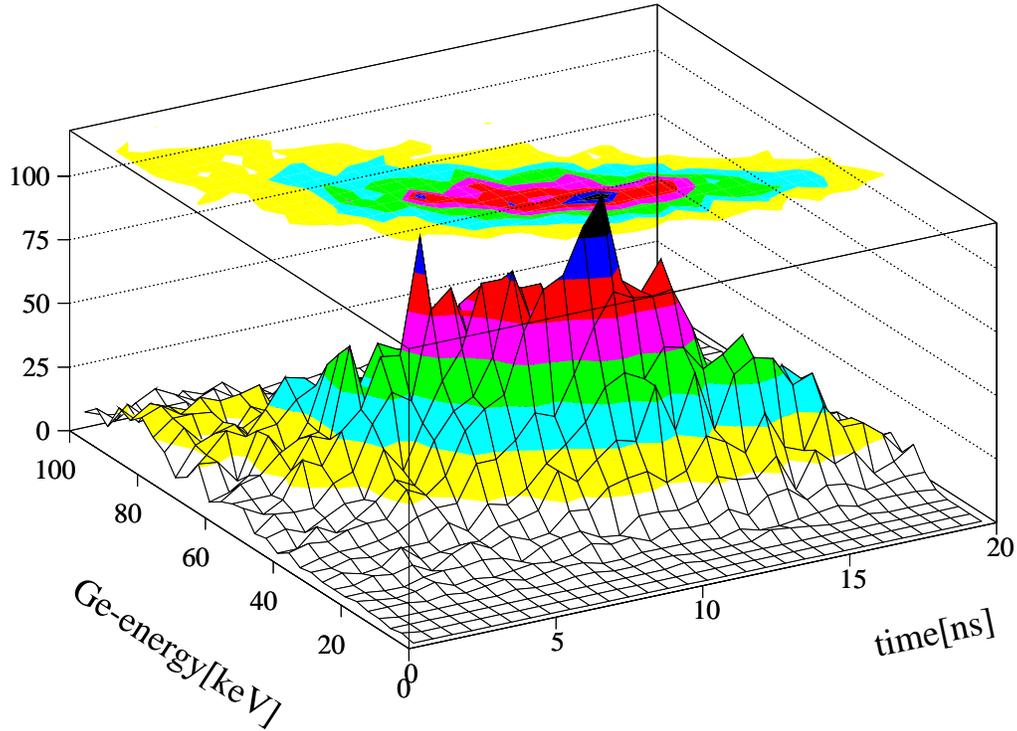}
\caption{Neutron flight time and ionization signal of the HPGe of
coincident events. The events from elastic scattered neutrons lie in
the curved band, which can be seen in the projection on top of the plot.}\label{etof}
\end{figure}

The Germanium detector was calibrated with a $^{228}$Th source, the
TOF measurements with several delays from 2 ns up to 150 ns. The
energy resolution of the Germanium detector was 1.55 keV at 80 keV. The
time resolution of the scintillators was $\le$ 1.5 ns. In Fig.~
\ref{etof}
the
energy deposited inside the germanium detector is plotted as function of
the TOF of neutrons detected in one neutron detector. 
The events from elastically scattered
neutrons form a continuous band towards the lower right. 

From this measurement the ionization efficiency of germanium atoms in
germanium can be calculated \cite{bau97a}:

\be
E_R = E_n\frac{2 m_{Ge} m_n +  m_n^2 - m_n^2 cos2\phi +
m_n cos\phi\sqrt{2 (2m_{Ge}^2-m_n^2 + m_n^2
cos2\phi)}}{(m_{Ge}+m_n)^2}
\label{gl9}
\ee

\ni Here E$_R$ is the energy of the recoil nucleus, E$_n$ is the
incident neutron energy, m$_{n}$ and m$_{Ge}$ are masses of neutron
and germanium nuclei, $\phi$ is the laboratory scattering angle of the
neutron.  The neutron energy E$_{n}$ as function of the flight time and energy
loss of the neutron which is equal to the recoil energy of the
germanium nucleus can be calculated from:

\be{}
t_n=d_1\sqrt{\,\frac{m_n}{2\,E_n}} + d_2\sqrt{\,\frac{m_n}{2(E_n-E_R)}},
\label{gl14}
\ee{}

\begin{figure}[h]
\hspace*{-1cm}
\epsfxsize=12cm
\hspace*{2.5cm}
\epsffile{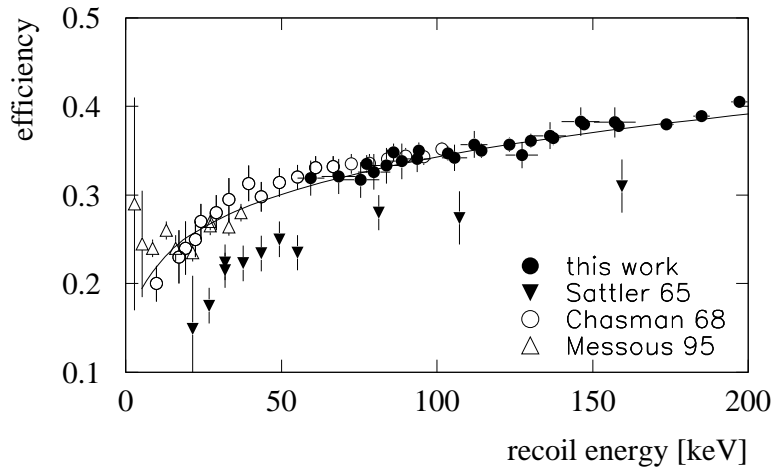}
\caption{Ionization efficiency as function of 
recoil energy. This measurement is marked by solid circles,
open circles mark \cite{cha65,cha68}, solid triangles 
 \cite{sat65} and open triangles \cite{mes95}. 
For comparison the calculation from Lindhard
\cite{lin63} is shown. Not shown are the results from \cite{jon75,jon71}
for recoil energies between 0.26keV and 1.75keV. }\label{eff}
\end{figure}

\ni where d$_{1}$ is the distance from the copper target to the germanium
detector; d$_{2}$ from germanium detector to the neutron detector.
Taking both formulae the recoil energy can be calculated and compared to
the ionization energy measured by the germanium detector. The
ionization efficiency is given by the ratio of ionization energy to
recoil energy. The results are plotted in figure \ref{eff}. The experimental
values are in good agreement with the theory of Lindhard which is also
verified by other experiments \cite{sat65,cha65,cha68,mes95,jon71,jon75}  down to 0.3 keV. Therefore we conclude that we
have measured recoil events from elastically scattered neutrons inside the
germanium detector. 

\begin{figure}[h]
\parbox{7cm}{\epsfxsize=7cm\epsffile{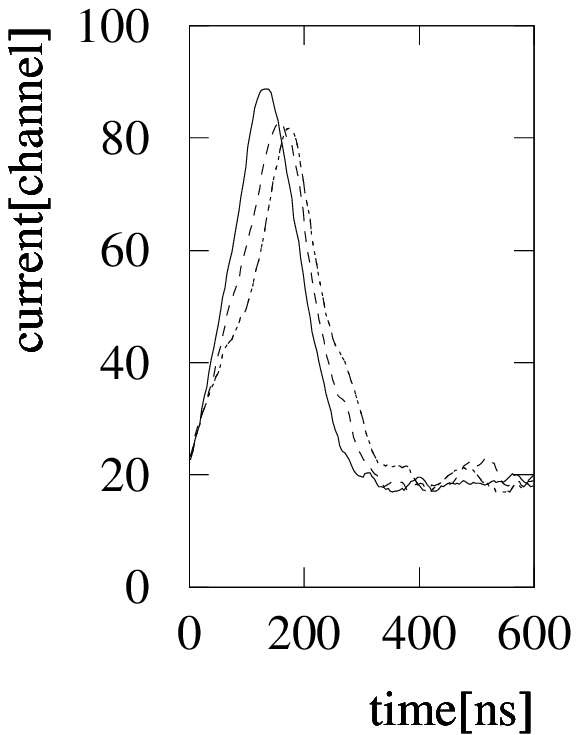}}
\hspace*{-2cm}\parbox{7cm}{\epsfxsize=7cm\epsffile{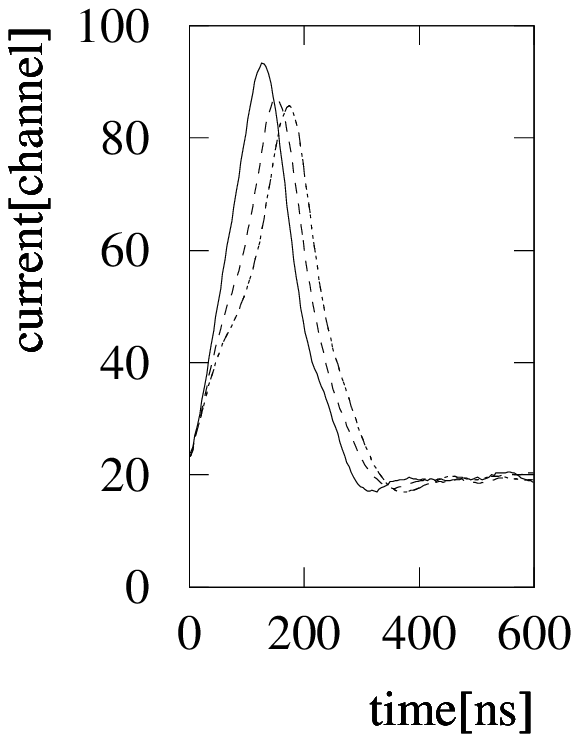}}
\caption{Mean pulse shapes of 80 keV neutron interactions with 88 ns, 112
ns and 136 ns rise time (left). Mean pulse shapes of 80 keV $\gamma$--interactions with 88 ns, 112
ns and 136 ns rise time (right). The pulses are differentiated with 50ns and
integrated with 20ns time constant. Both type of pulses are measured
with the n--type HPGe--detector.
}\label{ne}
\end{figure}

The pulse shapes of  recoil events were
obtained by differentiation of the customary integrating preamplifier output
with 20 ns time constant. The signal was also integrated with 20
ns. To reduce the noise level we selected the pulse shapes according
to their rise times and calculated mean pulse shapes. The mean pulse
shapes were calculated by adding the individual pulses of certain rise 
times (ca. 100 pulses per rise time) and dividing by the number of pulses. 
In Fig.~
\ref{ne} we show as an example mean pulse shapes for 80keV Ge recoil
nuclei from neutron scattering (left picture) with rise times of 88ns, 
112ns and 136ns.

\section{$\gamma$--Pulse shapes compared to nuclear recoils}
Production of nuclear recoil pulses in a low--level experiment for a
calibration measurement is not only a great effort but also polluting
the experiment, since the neutrons would activate the detector and
its shielding. Thus one has to think of a different source for
generating pulse
shapes to calibrate any pulse shape discrimination method. The usefulness of
$\gamma$--sources is obvious. We have sampled pulse shapes from a
$^{133}$Ba source by using the low energy Ba lines at 53keV and 80keV
and the $^{133}$Cs X--ray lines at 30keV and 35keV.

The pulse shapes in a coaxial detector depend on
the interaction radius. Therefore we have radiated the detector using a lead
collimator at different radial positions. The resulting rise times as
a function of the collimator position can be seen in Tab.~\ref{t2}.
The shape of the pulses depends on the motion of the charge carriers
in the electric field inside the Ge detector \cite{kno89}. 
Shown in the table are also the 
rise times corrected for the 
differentiation of the pulses with 50ns shaping time and the
rise times calculated 
under the assumption of a true coaxial
Ge--detector and a constant electron drift velocity of 10$^7$ cm/s
\cite{kno89}. 
For the n--type coaxial
Ge detector the collection time of the electrons (which are the majority
charge carriers and move towards the inner contact) dominates the time 
response of the detector \cite{kno89}.
%inner contact is the n$^+$--contact and the majority
%charge carriers are the electrons. 
Thus interactions
at smaller detector radii should show a smaller risetime than
interactions which take place in the outer part of the detector. This
dependence has been in principle confirmed in the behaviour of the pulses at low
energies. 
The difference between calculated and measured rise
times is due to the irradiation of the detector from the top, where the 
effect of the closed ended geometry is most visible. Since our aim was
to measure pulse shapes of different rise times and not to determine
the interaction radius from the measured rise time, this effect is of
no importance for this work.

In order to reduce the noise background the pulse shapes from
$\gamma$--ray events were selected and summed in the same way as the nuclear
recoil pulses described in Section 2. As example we show in Fig.~
\ref{ne} (right picture) 80keV pulse shapes from $\gamma$--ray
events. The three risetime classes are the same for nuclear recoil and
$\gamma$--ray events. The number of summed pulses was ca. one thousand
per risetime.

There is obviously no difference between nuclear recoil and $\gamma$--ray
pulses within the timing resolution of Ge--detectors. 
The mean $\gamma$--pulses are smoother because of the higher
statistics of the accumulated data with the $\gamma$--ray source.
We conclude that it is not possible to differentiate between
$\gamma$--ray-- and Ge--recoil--events by means of the pulse shape.

Thus a relative
background suppression method based on pulse shape analysis like for NaI
scintillators \cite{spo94,ber96} will not be applicable for
Ge--detectors. Consequently, one can calibrate the pulse shape
analysis for nuclear recoils by $\gamma$--ray sources.
This is an 
easy to handle method which needs no sophisticated
experiments and without the risk of activating the detector 
and radiation damage. 

\section{Comparison of pulse shapes from microphonic events with
  $\gamma$--interaction pulses }

When placed in a low level environment 
Germanium detectors are very sensitive to microphonic noise.
Microphonism constitutes one of the main limitations of Germanium detectors
in the low energy region and rends the evaluation of low energy
spectra ambiguous. The usual way of discriminating against
microphonism is to to use the timing information of each  event
\cite{bec94}. The time distribution of all events in the spectrum is
computed and cuts are set on the number of events per a certain time
interval. This method makes the assumption that microphonic events
occur in bursts and leads to run time losses up to
40\%. A method of analysing the pulse shape of each individual event
would be much more efficient.

%To record a library
%of typical microphonic pulses we used one of the enriched detectors
%from the Heidelberg--Moscow--Experiment \cite{hei97} situated in the
%Gran Sasso Underground Laboratory.
%It is a p--type enriched $^{76}$Ge--detector with an active mass of
%2.76 kg in a low level cryostat with 60cm distance between 
%FET and preamplifier. To reach a low energy threshold and record the pulse
%shape of each event we built up an  electronic and trigger
%system separate from the Heidelberg--Moscow--Experiment. The
%preamplifier energy 

To record a library
of typical microphonic pulses we used the small p--type natural  Ge--detector
of the HDMS--Experiment \cite{bau97} situated in the
Gran Sasso Underground Laboratory.
The detector has an active mass of
202 g and %an energy threshold of 2.5 keV. 
is situated in a low level cryostat with 60cm distance between 
FET and preamplifier. To reach a low energy threshold and record the pulse
shape of each event we built up a special  electronic and trigger
system. The
preamplifier energy 
signal is divided, amplified with 2$\mu$s  and 4$\mu$s
shaping time and measured by 13bit ADCs. The ADCs deliver fast, so
called peak--detect signals, which are subsequently used for
trigger purposes.  
The faster 2$\mu$s shaped signal serves as a stop
signal for the 250MHz flash ADC  which records the pulse
shapes. However, the 2$\mu$s shaped signal  yields a worse energy resolution and
thus a higher energy threshold because of the remaining higher noise
level of the baseline. 
The best energy resolution (1.87 keV at 1332keV) and threshold
(2.5keV) are obtained using the 4$\mu$s shaping. 
The preamplifier's timing output  is divided into four branches then
differentiated, integrated and amplified in timing filter amplifiers
with different time constants. The differentiation and integration
time constants are (50ns, 50ns), (100ns, 100ns), 
and (200ns, 200ns). The signals are amplified in two different 
ways to record
both low and high energetic pulses. For the purposes of this paper
the (50ns, 50ns) shaped pulses are most suitable.
The obtained pulse shapes are recorded 
with flash ADCs.   

\begin{figure}[h]
%\hspace*{1cm}
\parbox{10cm}{\epsfxsize=14cm\epsffile{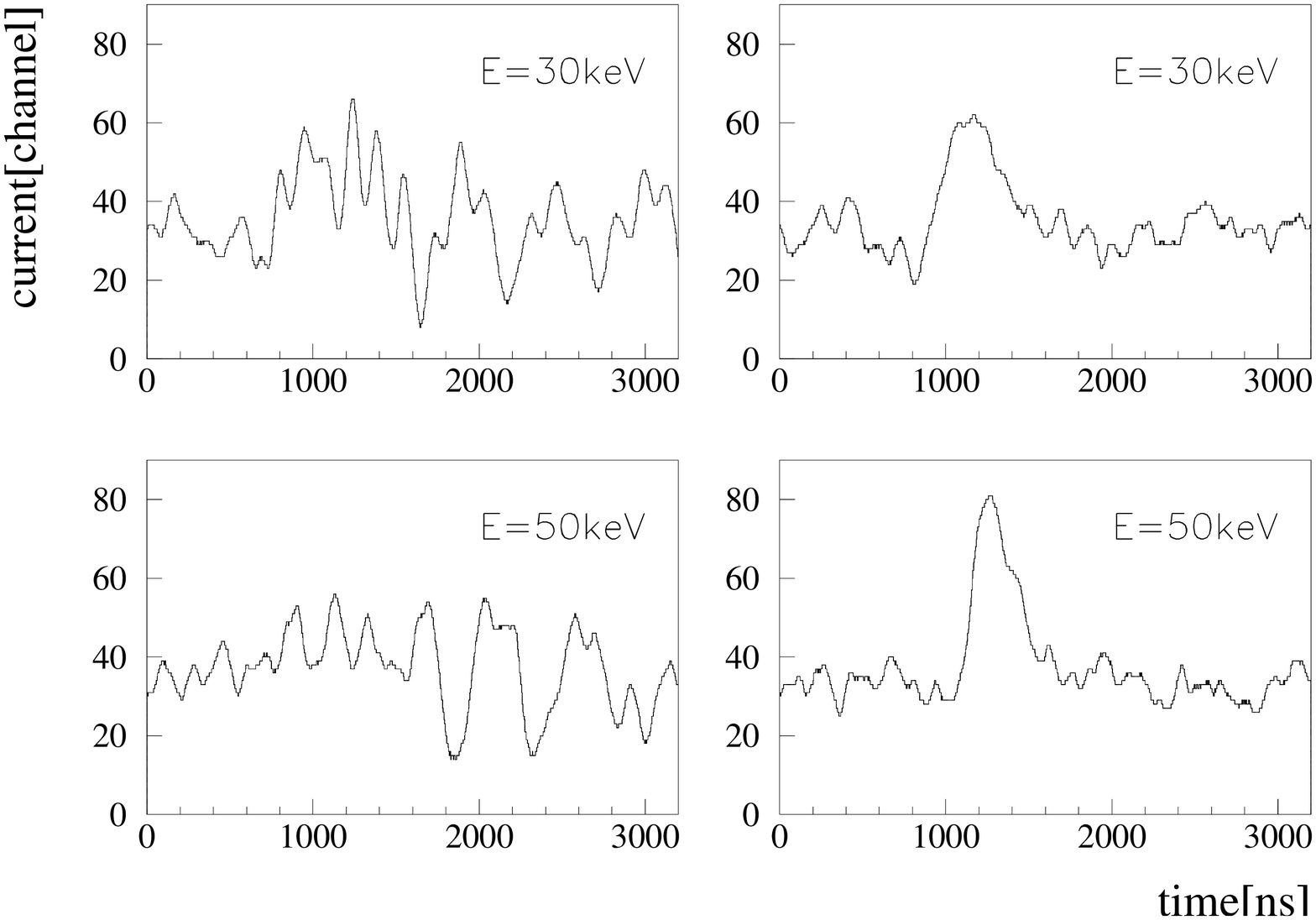}}
%\hspace*{1cm}\parbox{7cm}{\epsfxsize=10cm\epsffile{mikro2.eps}}
\caption{Individual pulse shapes of microphonic--events (left) and
  $\gamma$--events (right) in the low level p--type 
 Ge--detector. The pulses are differentiated with 50ns and
integrated with 50ns time constant. }\label{inner_pulse}
\end{figure}

We recorded microphonic pulse shapes with energies up to 60keV. 
$\gamma$--ray pulses with comparable energies were measured with an
EuTh source. 
Fig. ~\ref{inner_pulse} shows examples of both types
of pulse shapes with the same energies. On the left side are
microphonic pulses, $\gamma$--ray pulses are on the right side.
The patterns of the two kind of pulses are clearly different
and it is obvious that microphonic pulses are not like the baseline
noise  which is present in the microphonic free $\gamma$--ray pulses.
For a more quantitative comparison  between  microphonic-- and
$\gamma$--ray pulses we  suggest several
discrimination methods. One might analyse the power spectra of the pulses, compute
the second derivative and count the number of extrema 
or compute the integrated signal. Which, or which combination of the
above 
mentioned methods will deliver the highest rejection efficiency and
will be applied  
has yet
to be seen. 
Moreover the characteristics of microphonic pulses will depend on the
single detector and its operational environment. Thus it is not
reasonable to further investigate the different methods in this paper.

\section{Conclusion and Outlook}
Pulse shapes of recoil events from neutron scattering inside a
germanium detector were collected 
for the first
time. The measured ionization
efficiencies of recoiling germanium atoms in germanium 
are in good agreement with the theory of Lindhard and
earlier measurements.
We use this confirmation as cross check for our sample of nuclear
recoil pulse shapes. A difference to the pulse shapes from $\gamma$--ray
interaction has not been found.
For a calibration of
the nuclear recoil pulse shape we confirm the reasonable practice  to
use $\gamma$--ray
 sources instead of neutrons as a calibration standard.
We found instead a relevant difference between nuclear recoil
pulse shapes ( $\gamma$--pulse shapes) and microphonic pulses.
 Thus further development of an electronic
noise reduction method for dark matter experiments like \cite{bau97}
is possible by
measuring the pulse shape of each recorded event. 
This microphonic
noise mainly obscures energy spectra for WIMP detection in the most
interesting near--threshold energy region. A discrimination
method against microphonics would eliminate one of the last systematic
uncertainties for Ge--detectors. We are confident to apply such a method
in our new dark matter experiment, starting it's operation during this year, the 
Heidelberg Dark Matter Search (HDMS) Experiment \cite{bau97}.

\bibliographystyle{unsrt}

\begin{thebibliography}{10}


\bibitem{kam96}G. Jungmann, M. Kamionkowski and K. Griest,
Phys. Rep. 267 (1996) 195.

\bibitem{smi90}P.~F. Smith and J.~D. Lewin, Phys. Rep. {\bf 187} (1990) 203.


\bibitem{spo94}
N.~J. Spooner and P.~F. Smith,
\newblock {\em Phys. Lett.}, B 314:430, 1994.


\bibitem{shu92}
T.~Shutt et~al.
\newblock {\em Phys. Rev. Lett.}, 69:3425, 1992.

\bibitem{garcia95}
E. Garcia et~al.,
\newblock {\em Phys. Rev.}, D 51:1458, 1995.

\bibitem{bec94}
M.~Beck et~al., Heidelberg--Moscow Collaboration, 
\newblock {\em Phys. Lett.}, B 336:141--146, 1994.

\bibitem{ber96}
R.~Bernabei et~al.
\newblock {\em Phys. Lett.}, B 389:757, 1996.


\bibitem{bau97a}
L.~Baudis .
\newblock {\em Diploma Thesis}, 1997, unpublished.


\bibitem{hei97}
L.~Baudis et~al., Heidelberg--Moscow Collaboration, 
\newblock {\em Phys. Lett.}, B 407:219, 1997

\bibitem{bam91} 
G. J. Bamford, A. C. Rester, R. L. Coldwell, C. M. Castanada 
\newblock IEEE Trans. Nucl. Sci. NS-38 (1992) 200.

\bibitem{asp94} 
B. Aspacher and A. C. Rester, 
\newblock {\em Nucl. Inst. and Meth.} A 338 (1994) 516.

\bibitem {rot84} 
J. Roth, J. H. Primbsch and R. P. Lin, 
\newblock IEEE Trans. Nucl. Sci NS-31 (1984) 367.
%\bibitem{bau97b}
%L.~Baudis et~al.
%\newblock {\em Nucl. Inst. and Meth.}, A 358:265--267, 1997.


\bibitem{sat65}
A.~R. Sattler.
\newblock {\em Phys. Rev.}, 138:A1815, 1965.

\bibitem{cha65}
C.~Chasman, K.~W. Jones, and R.~A. Ristinen.
\newblock {\em Phys. Rev. Lett.}, 15:245 (E) 684, 1965.

\bibitem{cha68}
C.~Chasman, K.~W. Jones, and W.~Kraner, H. W.and~Brandt.
\newblock {\em Phys. Rev. Lett.}, 21:1430, 1968.

\bibitem{mes95}
Y.~Messous et~al.
\newblock {\em Astroparticle Physics}, 3:361, 1995.

\bibitem{ham86}
J.~W. Hammer et~al.
\newblock {\em Nucl. Inst. and Meth.}, A 244:455--476, 1986.

\bibitem{lin61}
J.~Lindhard and M.~Scharff.
\newblock {\em Phys. Rev.}, 124:128, 1961.

\bibitem{jon75}
K.~W. Jones and H.~W. Kraner
\newblock {\em Phys. Rev.}, {\bf A 11 } (1975) 1347--1353.

\bibitem{jon71}
K.~W. Jones and H.~W. Kraner
\newblock {\em Phys. Rev.}, {\bf C 4 } (1971) 125--129.

\bibitem{esc94}
M.~Eschenauer et~al., Nucl. Inst. and Meth. {\bf A 340} (1994) 364--370.

\bibitem{kra80}
H.~W. Kraner, IEEE Trans. Nucl. Sci {\bf NS-27} (1980) {218}.

\bibitem{lin63}
J.~Lindhard, V.~Nielsen, M.~Scharff, and P.~V. Thomsen, Mat. Fys. Medd. Dan.
  Vid. Selsk. {\bf 33} (1963) 1.

\bibitem{kno89}
G.~F. Knoll, {\em {Radiation Detection and Measurement}}, 2. ed., Wiley, 1989.


\bibitem{bau97}
L.~Baudis et~al.
\newblock {\em Nucl. Inst. and Meth.}, A 358:265--267, 1997.


\end{thebibliography}

\newpage

\begin{table}[h]
{\bf 
\caption{Maximal neutron energies for E$_{p}$ = 16 MeV under 30$^{\circ}$.}
\label{t1}
\vskip0.3cm
}
\centering

\begin{tabular}{lc}
\hline
Energy [MeV] & Reaction\\
\hline
13.77 & p(${^7}$Li,$^7$Be)n \\
13.75 & p($^{65}$Cu,$^{65}$Zn)n \\   
11.78 & p($^{63}$Cu,$^{63}$Zn)n \\
\hline
\end{tabular}
\end{table}

\begin{table}
\caption{Tabulation of the $\gamma$--pulse shape rise times (t$_r$) as
function of the Ge--detector radius. Shown are the measured (10--90\%)
rise times, the (0--100\%) rise times calculated from the measured (10--90\%)
rise times, the rise times corrected  for the
differentiation with $\tau$=50ns and the rise times calculated under
the assumption of a true coaxial geometry and a constant electron
drift velocity of 10$^7$ cm/s \cite{kno89}.   }
\label{t2}
\vskip0.3cm
\begin{center}
\begin{tabular}{ccccc}
\hline
radial position $[$cm$]$ & t$_r$(10--90\%) $[$ns$]$ & t$_r$
(0--100\%)&  $t_{r,corrected}$ $[$ns$]$  & $t_{r, calculated}$ $[$ns$]$  \\
\hline
0.0$\pm$0.1 & 60$\pm$20 & 72$\pm$20 & - &  -\\
0.5$\pm$0.1 & 60$\pm$20& 72$\pm$20& 40$\pm$11 & 10 \\
1.0$\pm$0.1 & 120$\pm$20& 144$\pm$20& 134$\pm$18 & 60\\
1.5$\pm$0.1 & 150$\pm$20& 180$\pm$20 & 173$\pm$19 & 110\\
2.0$\pm$0.1 & 160$\pm$15& 192$\pm$15& 187$\pm$15 & 160\\
2.5$\pm$0.1 & 180$\pm$10& 216$\pm$10& 213$\pm$10 & 210\\ 
\hline
\end{tabular}
\end{center}
\end{table}

\end{document}